\documentclass[preprint,12pt]{aastex}
\usepackage[caption=false]{subfig}
\usepackage{graphicx}

\begin{document}
\pagestyle{plain}
\pagenumbering{arabic}

\begin{center} 
%\bfseries\uppercase{%
{\large \textbf {
A Wideband Polarization 
Survey of the Extragalactic Sky at 2-4 GHz: A Science White Paper for the VLA Sky Survey
}}
\end{center}

\noindent{\footnotesize Sui Ann Mao (NRAO, U Wisconsin Madison, mao@astro.wisc.edu),
Julie Banfield (CASS, CSIRO),
Bryan Gaensler (CAASTRO, University of Sydney),
Lawrence Rudnick (U Minnesota),
Jeroen Stil (U Calgary),
Cormac Purcell  (University of Sydney),
Rainer Beck (MPIfR),
Jamie Farnes (University of Sydney),
Shane O'Sullivan (University of Sydney),
Dominic Schnitzeler (MPIfR),
Tony Willis (DRAO),
Xiaohui Sun (University of Sydney),
Ettore Carretti (CASS, CSRIO),
Klaus Dolag (U Munich),
Dmitry Sokoloff (Moscow State University),
Roland Kothes (DRAO),
Maik Wolleben (U Calgary),
George Heald (ASTRON),
Joern Geisbuesch (DRAO),
Tim Robishaw (DRAO),
Jose Afonso (Observatorio Astronomico de Lisboa),
Antonio Mario Magalh\~aes (U S\~ao Paulo),
Britt Lundgren (U Wisconsin Madison),
Marijke Haverkorn (Radboud University Nijmegen),
Niels Oppermann (CITA),
Russ Taylor (U Calgary)
}

%\begin{abstract}
\begin{center}
{ \textbf {Abstract}
}
\end{center}
A VLA Sky Survey of the extragalactic sky at S band (2-4 GHz) with polarization information can uniquely probe the magneto-ionic medium in a wide range of astrophysical environments over cosmic time. For a shallow all-sky survey, we expect to detect over 4 million sources in total intensity $>$ 0.45 mJy beam$^{-1}$ and over 2.2$\times$10$^5$ sources in polarized intensity. With these new observations, we expect to discover new classes of polarized radio sources in very turbulent astrophysical environments and those with extreme values of Faraday depth. Moreover, by determining reliable Faraday depths and by modeling depolarization effects, we can derive properties of the magneto-ionic medium associated with AGNs, absorption line systems and galaxies, addressing the following unresolved questions: (1) What is the covering fraction, the degree of turbulence and the origin of absorption line systems? (2) What is the thermal content in AGNs and radio galaxies? (3) How do AGNs and galaxies evolve over cosmic time? (4) What causes the increase in percentage polarization with decreasing flux densities at the low flux density end of the polarized source count? (5) What is the growth rate of large-scale magnetic fields in galaxies? 
%\end{abstract}

\clearpage
\section{Introduction}
\subsection{Faraday Rotation Effect as a Probe of the Magneto-ionic Medium} 
When linearly polarized light travels through a magneto-ionic medium, the Faraday rotation effect modifies the complex polarization of the incident radiation. The Faraday depth $\phi$ (rad~m$^{-2}$), given by  
%\begin{equation}
$\phi =0.812 \int ^{observer}_{source} {n_{e}(l){B_{\parallel} (l)}} dl$
%\label{vlaeq:rmdef}
%\end{equation}
is used to describe the amount of Faraday rotation along a particular sightline. It provides a measure of the thermal electron content  $n_e(l)$ (in cm$^{-3}$) and magnetic fields $B_{\parallel}(l)$ (in $\mu$G) along the line of sight. Faraday depth $\phi$ equals rotation measure (RM) in the simplest case that the background radiation is Faraday rotated in an external uniform foreground medium. 

Wideband receivers of modern radio telescopes allow one to implement the RM synthesis technique (e.g., Burn 1966, Brentjens \& de Bruyn 2005) to extract polarized emission at different Faraday depths. It has been widely used to extract Faraday depths of both compact point sources (e.g., Mao et al. 2010) and diffuse polarized emission (e.g. Schnitzeler et al. 2009, Heald et al. 2009). However, Farnsworth et al. (2011) show that relying on RM synthesis alone could lead to misinterpretation of the underlying Faraday structure, especially in the presence of multiple RM components. The authors suggest that fitting models directly to the fractional $q$ and $u$ as a function of $\lambda^2$ is the only reliable way to determine the true Faraday structure along the line of sight (see also e.g., O'Sullivan et al. 2012). %RM determination data challenge led by 

\subsection{The Need for a Wideband Polarization Survey} 
Polarized radiation passing through a magneto-ionic medium is subjected to various external or internal depolarization processes (e.g., Burn 1966, Tribble 1991, Sokoloff 1998) which can result in very similar behaviors of $Q/I(\lambda^2)$ and $U/I(\lambda^2)$ within a narrow frequency range. O'Sullivan et al. (2012) have shown that wideband polarimetric observations are necessary to reveal the underlying depolarization effect. In Figure 1, we use an example taken from O'Sullivan et al. (2012) to illustrate this. Polarization measurements of an extragalactic source between 1-3 GHz have been fitted to two different Faraday structures. With only a narrow frequency coverage (for example, between 1.67-1.87 GHz), one would not be able to distinguish between the two Faraday structures /depolarization models. {\bf Therefore, degeneracies between different types of depolarization behavior, and hence the underlying properties of the line-of-sight magneto-ionic medium, can only be broken by wideband spectro-polarimetry.} The WIDAR correlator on the upgraded Jansky VLA has opened a brand new window for wideband polarization work, enabling us to characterize properties of the magneto-ionic medium in AGNs and in galaxies across a wide range of redshifts, as will be outlined in this white paper. 
%- a bit too obvious.  change that..

%While it is challenging to distinguish between the two models in the presence of noise 0.08$\ge$$\lambda^2$$\ge$0.045 cm$^2$, it becomes feasible when a larger bandwidth is considered. 
\begin{figure}[h]
\centering
\epsscale{0.85}
\plotone{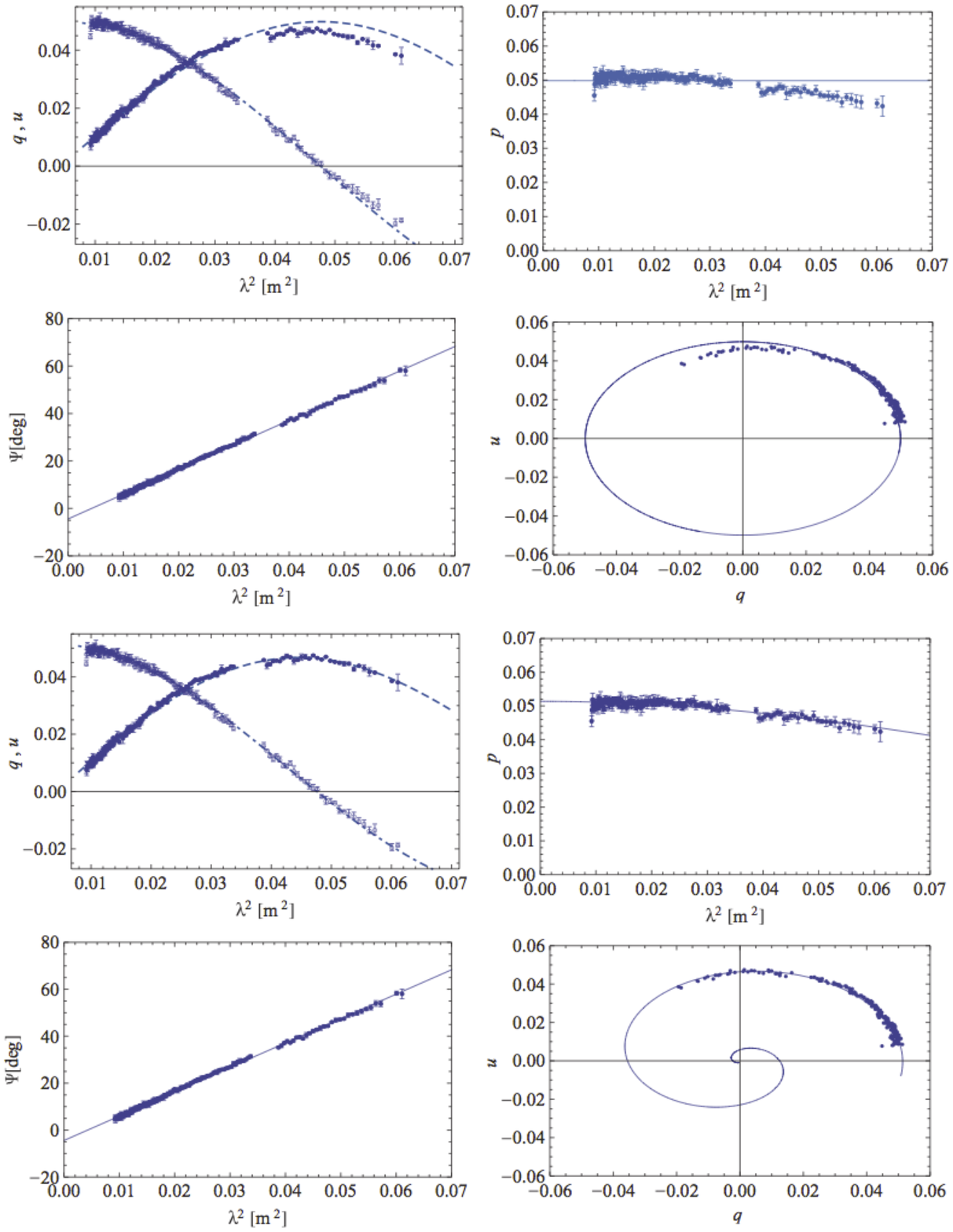}
\caption{Plots of the fractional $q$($\lambda^2$) and $u$($\lambda^2$), polarized fraction as a function of $\lambda^2$, polarization angle as function of $\lambda^2$, and the $q$ $u$ track for an extragalactic source in O'Sullivan et al. (2012). The top 4 panels show the data fitted to a single Faraday rotation component, and the bottom 4 panels show the data fitted to a single component plus additional depolarization due to external Faraday dispersion. The model fits are denoted by solid lines in the plots.}

\end{figure}

\section{The Proposed Survey Setup} 
%{\bf We propose a polarization survey to be conducted at S band (2-4 GHz) with C Array configuration (7" resolution)

\subsection{Frequency band}
We argue that an 2-4 GHz (S band) polarization survey will facilitate discoveries of new populations of extragalactic sources: those depolarized at low frequencies; those with extended structures in Faraday depth space, and those with extremely large Faraday depths. S band polarization data from the VLASS, supplemented by L band polarimetric surveys to be carried out with SKA pathfinders: the POSSUM (POlarisation Sky Survey of the Universe's Magnetism, Gaensler et al. 2010) with the ASKAP (Australia SKA Pathfinder) (1130-1430MHz, ASKAP-12 early science may potentially cover the full ASKAP band 700-1800 MHz), the WODAN (Westerbork Observations of the Deep APERTIF Northern-Sky, R\"ottgering et al. 2010) ($\Delta\nu$=300 MHz between 1000-1750MHz), and the FRIGG (Faraday Rotation Investigation of Galaxies and Groups, PI: Beck) at the APERTIF, will form an extremely powerful polarization data set of extragalactic sources.
% Mention of SPASS ATCA followups, COMOS JVLA and PIGGS %This combined data set would yield nearly continuous frequency coverage from 1-4 GHz at a density of $\sim$ 100 polarized sources deg$^{-2}$. ??

\begin{itemize}
\item {\bf Detect populations of sources that are depolarized at frequencies $<$ 2 GHz: }
%FR not much at even higher freq
The main advantage of observing at 2-4 GHz compared to lower frequencies is the smaller wavelength-dependent depolarization effects which can be quantified by the dispersion of intrinsic rotation measure ($\sigma_{RM}$) within the beam volume. Extragalactic polarized sources can have $\sigma_{RM}$ up to few 10s of rad m$^{-2}$ across the source (e.g., Goodlet et al. 2005). If these sources are unresolved, Figure 1 in Arshakian \& Beck (2011) shows that their optimal observing frequency would be in the 2-4 GHz range. As an example, if the Burn (1966) external Faraday dispersion law applies, the fractional polarization would increase more than an order of magnitude when observing at 3 GHz rather than at 1.5 GHz for $\sigma_{RM}$$\sim$ 30 rad m$^{-2}$. Hence, a 2-4 GHz polarization survey gives us access to a different class of targets with large $\sigma_{RM}$. Extremely turbulent astrophysical environments with large fluctuation in gas densities and magnetic fields such as inner regions of jets, cores of massive galaxy clusters \footnotemark[1]\footnotetext[1]{We do not include polarization science on clusters in this white paper: there is a separate white paper (Tracy Clarke et al.) on cluster science which has a polarization section.}and starburst galaxies could potentially be polarized at frequencies $>$ 2 GHz.  

\item {\bf Detect Extended Structures in Faraday Depth Space}: When synchrotron emitting and Faraday rotating material is mixed in, it produces an extended structure in the Faraday spectrum. The scale at which the sensitivity to extended structures drops to 50\% is given by $\pi/\lambda_{\rm min}^2$. By observing at 2-4 GHz, we greatly enhance the detectability of these extended structures: the largest detectable scale is $\sim$ 560 rad m$^{-2}$, a factor of 4 larger than that at 1-2 GHz. %(add example)

\item {\bf Improvement of the Resolution in Faraday Depth Space}: The precision of determining Faraday depths depends on ${2\sqrt{3}}/({\lambda_{max}^2-\lambda_{min}^2})$. With 2-4 GHz data alone, the resolution in Faraday depth space is $\sim$ 200 rad m$^{-2}$, which is $\sim$ two times worse than that measured at L band at the Jansky VLA\footnotemark[2]\footnotetext[2]{The ideal L band resolution in Faraday depth space is 4 times smaller than that at S band. In reality, RFI leads to incomplete wavelength coverage between 1-2 GHz and it results in a Faraday depth resolution of $\sim$ 90 rad m$^{-2}$ instead.}(Mao et al. in prep). However, when supplemented with L band polarization data from WODAN and POSSUM, the resulting Faraday depth resolution will be improved to $\sim$ 40 rad m$^{-2}$.

%Though in reality, the resolution in Faraday depth space using the Jansky VLA at L band is more like $\sim$ 90 rad m$^{-2}. However, when combined with L band polarization information provided by POSSUM and WODAN, the 
%S band only 205 
%L band only  51.32 ideal , reality up to 90 rad m-2 (Mao et al. in prep)
%S+L band => 41 rad/m^2

\item {\bf Sensitivity to Larger Faraday Depths}: The maximum Faraday depth detectable in an observation relates to the channel width in wavelength squared space $\delta\lambda$ by ${\sqrt{3}}/{\delta\lambda^2}$. Consider fixed channel width of 1 MHz, 2-4 GHz observations will be sensitive to Faraday depths up to 5$\times$10$^5$  rad m$^{-2}$, a factor of 8 larger than the corresponding value for 1-2 GHz observations (assuming ideal wavelength coverage). A 2-4 GHz polarization survey will increase our potential to detect sources with extremely large Faraday depths, possibly associated with low-luminosity or core-dominated AGNs (e.g., Bower et al. 2002). 

\end{itemize}

Faraday depths can be readily extracted from 2-4 GHz  polarization observations from the Jansky VLA. In Figure 2, we show a direct fit to  $Q/I(\lambda^2)$ $U/I(\lambda^2$) of an extragalactic source consists of two RM components from recent 2-4 GHz Jansky VLA observations (Williams, Mao et al. in prep). 

\begin{figure}[h]
%\centering
\plotone{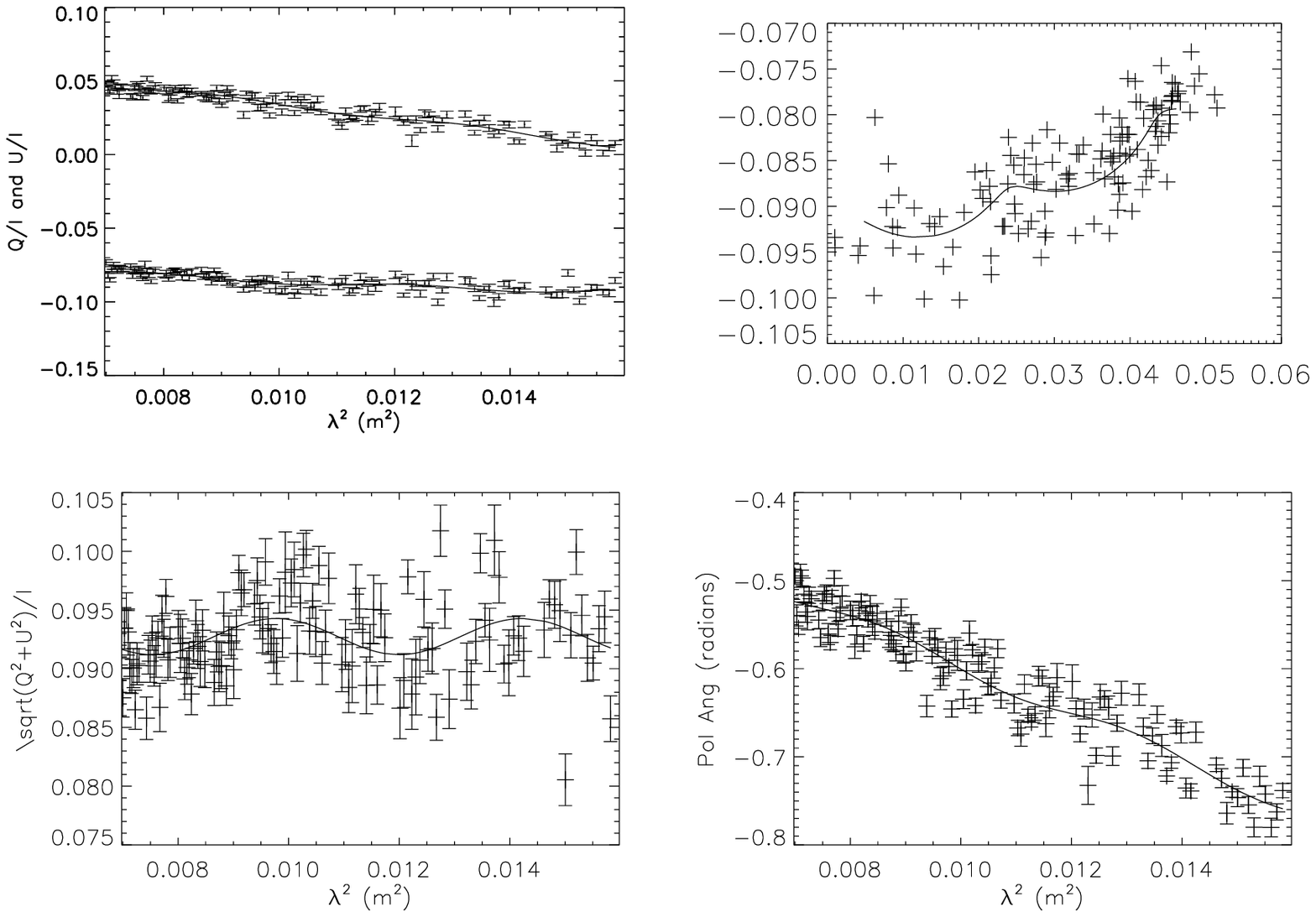}
\caption{Plots of the fractional $q$($\lambda^2$) and $u$($\lambda^2$), the $q$ $u$ track, polarized fraction as a function of $\lambda^2$, and the polarization angle as function of $\lambda^2$ for an extragalactic source from recent 2-4 GHz observations done at the Jansky VLA (Williams, Mao et al. in prep). This source can be well fitted by two Faraday-thin components.}
%Plots of the percent polarization (top), fractional Stokes $Q$ (middle) and  $U$ (bottom) as a function of $\lambda^2$ for two different sight lines effected by different depolarization effects. Red solid lines depict the depolarization trend due to a uniform magnetic field in a region where synchrotron emitting and Faraday rotating materials are well mixed. Black solid line the depolarization due to a foreground turbulent screen, external to the synchrotron emitting material.}
\end{figure}

\subsection{Array Configuration}
We choose C array configuration (7" resolution) for a good trade-off between surface brightness sensitivity and angular resolution, while minimizing effects from beam depolarization due to the turbulent nature of the magneto-ionic medium (Gaensler et al. 2005, Stil et al. 2007). The beam size of $\sim$ 7" can provide accurate astrometry for positional matching with other radio surveys and ancillary data in optical and IR wavelengths. The C array beam also roughly matches that of the WODAN and POSSUM surveys ($\sim$ 10"), which facilitates a joint analysis.

\subsection{Survey Depth and Coverage}

We will use a hexagonal grid of pointings to achieve uniform sensitivity across large regions in the sky. For a field of view of 15' at S band, we require $\sim$ 25 pointings per square degree. We assume that 3,000 hours of on-source time will be dedicated to the VLA Sky Survey (within the listed reasonable range on the VLASS faq page). Since the survey coverage is likely to be driven by the continuum science case, we consider the following four possibilities (They are summarized in Table 1): 
\begin{enumerate}
\item {\bf A shallow all-sky field}: covering the entire sky at $\delta$ $\ge$ $-$40$^\circ$ with uniform sensitivity.  This translates into spending $\sim$ 13 seconds per pointing. We will reach an rms of 90 $\mu$Jy beam$^{-1}$ and a 5$\sigma$ detection threshold of 0.45 mJy beam$^{-1}$. The all-sky coverage completely overlaps with the WODAN survey for $\delta$ $\ge$ $+$30$^\circ$and with the POSSUM survey for $-$40$^\circ$$\le$$\delta$ $\le$+30$^\circ$. %Therefore, for every sources detected at 2-4 GHz, there will be complementary 1-2 GHz data. %Based on theCompare the sensitivity with FIRST and NVSS, even a shallow all-sky field has a drastic improvement factor of X. 

\item {\bf A medium sensitivity field $\sim$ 2000 deg$^2$}: For example, covering $\sim$ 2,000 deg$^2$ in 3,000 hours translates into spending $\sim$ 3.3 mins per pointing. We will reach an rms of 22 $\mu$Jy beam$^{-1}$ and a 5$\sigma$ detection threshold of 0.11 mJy beam$^{-1}$. Substantially large regions around the North (South) Galactic Pole region should be covered to ensure overlap with WODAN (POSSUM). We can also cover a good fraction of the SDSS spectroscopic survey area. High Galactic latitude regions are desirable fields as the Milky Way's contribution to the observed Faraday depth towards extragalactic polarized sources and their scatter is small in general at large $|$b$|$. 

%Possible regions to be covered could include the North 

%and the South Galactic Poles which allow overlap with the SDSS (NGP) and both the WODAN and POSSUM surveys. We note also that the Milky Way's contribution to the observed Faraday depth towards extragalactic polarized sources and their scatter is minimized towards high Galactic latitudes.  

\item {\bf Deep fields covering several 100s deg$^2$}: For example, covering 500 deg$^2$ in 3,000 hours translates into $\sim$ 14.4 mins per pointing. We will reach an rms of 10.3 $\mu$Jy beam$^{-1}$ and a 5$\sigma$ detection threshold of 52 $\mu$Jy  beam$^{-1}$. The North and the South Galactic poles are possible regions to be covered, since they allow overlap with selected regions of the SDSS, and both the WODAN and POSSUM surveys. 

\item {\bf Ultra-deep fields covering several 10s deg$^2$}: For example, covering 50 deg$^2$  in 3,000 hours translates into $\sim$ 2.5 hours per pointing. We will achieve an rms of 3.5 $\mu$Jy beam$^{-1}$ and a 5$\sigma$ detection threshold of 17.5 $\mu$Jy  beam$^{-1}$. Fields with extensive multi-wavelength coverage and previous deep radio observations should be selected. Some possible fields include the Lockman hole field, the Bo\"otes field and the ELAIS-N field. % Great WALL?? 
 
\end{enumerate}

\begin{deluxetable}{lrrrrr}
%\setlength{\tabcolsep}{0.02in} 
%\rotate
\centering
\tabletypesize{\footnotesize}
\tablecolumns{7} 
\tablewidth{0pc} 
\tablecaption{Survey Depth and Coverage} 
\tablehead{   
\colhead {Survey Type} & \colhead{Area Covered} & \colhead{Sensitivity}   & \colhead{5$\sigma$ Detection Threshold} 
}
\startdata
Shallow & all sky $\delta$ $\ge$ $-$40$^\circ$   & 90 $\mu$Jy beam$^{-1}$  & 0.45 mJy beam$^{-1}$ \\
Medium & 2,000  deg$^2$ & 22 $\mu$Jy beam$^{-1}$  & 0.11 mJy beam$^{-1}$ \\
Deep & $\sim$several 100s deg$^2$  & 10.3 $\mu$Jy beam$^{-1}$ & 52 $\mu$Jy beam$^{-1}$ \\
Ultra-deep & $\sim$ several 10s  deg$^2$ & 3.5 $\mu$Jy beam$^{-1}$  &  18$\mu$Jy beam$^{-1}$ \\
 \enddata 
\label{surveysetup}
\end{deluxetable}

\subsection{Expected Source Counts and their Redshift Distributions}

In Figure 3, we show the expected redshift distribution of total intensity sources for four different possible types of survey: shallow all-sky (top panel), medium fields (2nd panel), deep fields (3rd panel) and ultra-deep fields (bottom panel) based on the SKADS simulations (Wilman et al. 2008, 2010). In Table 2, we list the expected number of sources observed in total intensity for the four different possible sky coverages. We note that for a shallow all-sky survey, the source distribution is mostly dominated by bright AGNs, therefore it may not be suitable for science goals targeting the low flux density end of the source count distribution. 
%A shallow survey will have enough re
%Also, a shallow survey may not be able to  provide enough redshift bins to study for example, the evolution of source properties as a function of cosmic time. no bright AGNs may have 

Since little is known about the polarized source count at 2-4 GHz, we use polarized source count at 1.4 GHz and a mean spectral index of extragalactic sources of -0.7 to place a lower limit on the polarized source density expected for the proposed VLASS. We used the polarized source count at 1.4 GHz derived from stacking the NVSS data (Stil et al., submitted) (Figure 4). We list the estimated number of polarized sources for the different types of surveys in Table 3.  A shallow all-sky survey will yield at least 2.2 $\times$ 10$^5$ polarized sources. To date, there has not been a blind large-area wide-band polarimetric survey done at 2-4 GHz with interferometers \footnotemark[3]\footnotetext[3]{The Jansky VLA COSMOS survey at S band (PI: Smol\v{c}i\'{c}) covers a small area of 2 square degrees, while the ATCA CABB observations at 1-3 GHz of polarized extragalactic sources (PI: Schnitzeler) are pointed observations to follow up sources discovered in the S-PASS survey (PI: Carretti).}. Therefore, the polarized source count measured from the proposed VLASS will provide important guidance to the survey design of the Square Kilometer Array at similar frequencies.

\begin{deluxetable}{lccccc}
\centering
\tabletypesize{\footnotesize}
\tablecolumns{7} 
\tablewidth{0pc} 
\tablecaption{Expected number of sources in total intensity estimated using the SKADS} 
\tablehead{   
\colhead {Survey type (sky coverage)} & \colhead{Normal Galaxies} & \colhead{Starbursts}   & \colhead{Radio Quiet AGNs} & \colhead{FRI} &\colhead{FRII}
}
\startdata
Shallow ($\sim$34,000 deg$^2$) & 3.0$\times$10$^5$ & 3.0$\times$10$^5$  & 2.3$\times$10$^5$ & 3$\times$10$^6$ & 1.8$\times$10$^5$\\
Medium ($\sim$2,000 deg$^2$) & 1.6$\times$10$^5$ & 1.4$\times$10$^5$  & 1.2$\times$10$^5$ & 4.9$\times$10$^5$ & 1.2$\times$10$^5$\\
Deep ($\sim$500 deg$^2$)& 1.3$\times$10$^5$ & 6.4$\times$10$^4$ & 6.9$\times$10$^4$ & 1.5$\times$10$^5$ & 2.2$\times$10$^3$\\
Ultra-deep ($\sim$50 deg$^2$)  & 8$\times$10$^4$ & 2.26$\times$10$^4$ & 3.5$\times$10$^4$ & 3.3$\times$10$^4$ & 328 \\
 \enddata 
\label{surveysetup}
\end{deluxetable}

\begin{deluxetable}{lcc}
\centering
\tabletypesize{\footnotesize}
\tablecolumns{7} 
\tablewidth{0pc} 
\tablecaption{Expected Number of Polarized Sources} 
\tablehead{   
\colhead {Survey Type} & \colhead{Source Count per sq deg (lower limit)} & \colhead{Total Source Count (lower limit)}  
}
\startdata
Shallow & 7 & 2.2$\times$10$^5$ \\
Medium  &19 & 3.8$\times$10$^4$\\
Deep& 34 & 1.7$\times$10$^4$ \\
Ultra-deep  & 76 & 3.8$\times$10$^3$ \\
 \enddata 
\end{deluxetable}

\begin{figure}[ht]

\begin{center}
\epsscale{0.45} 
\hspace*{-2cm}
\subfloat[Shallow all-sky]{\plotone{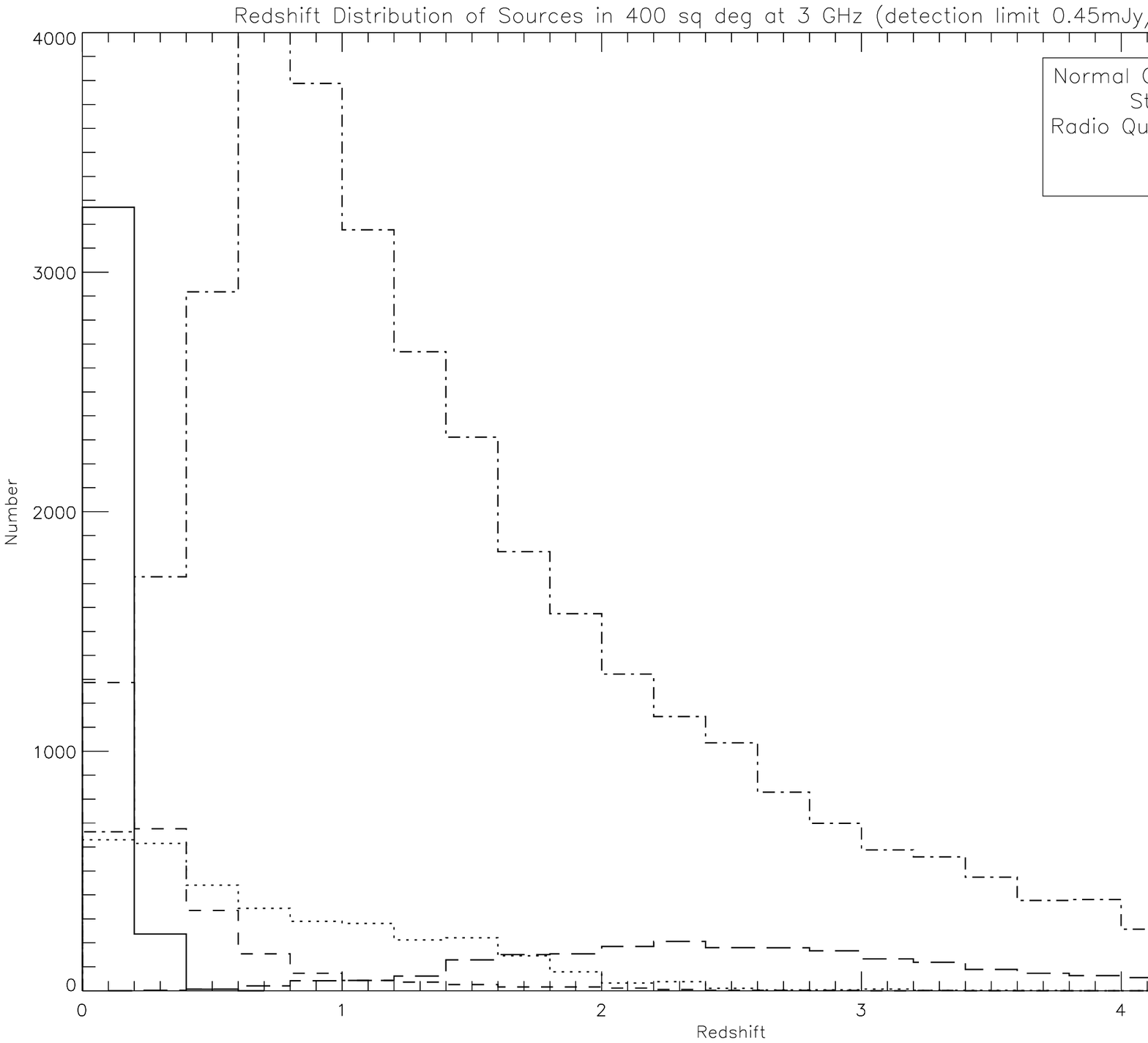}}%\includegraphics[width=3in]{combined_histo_shallow.eps}}
\subfloat[Medium]{\plotone{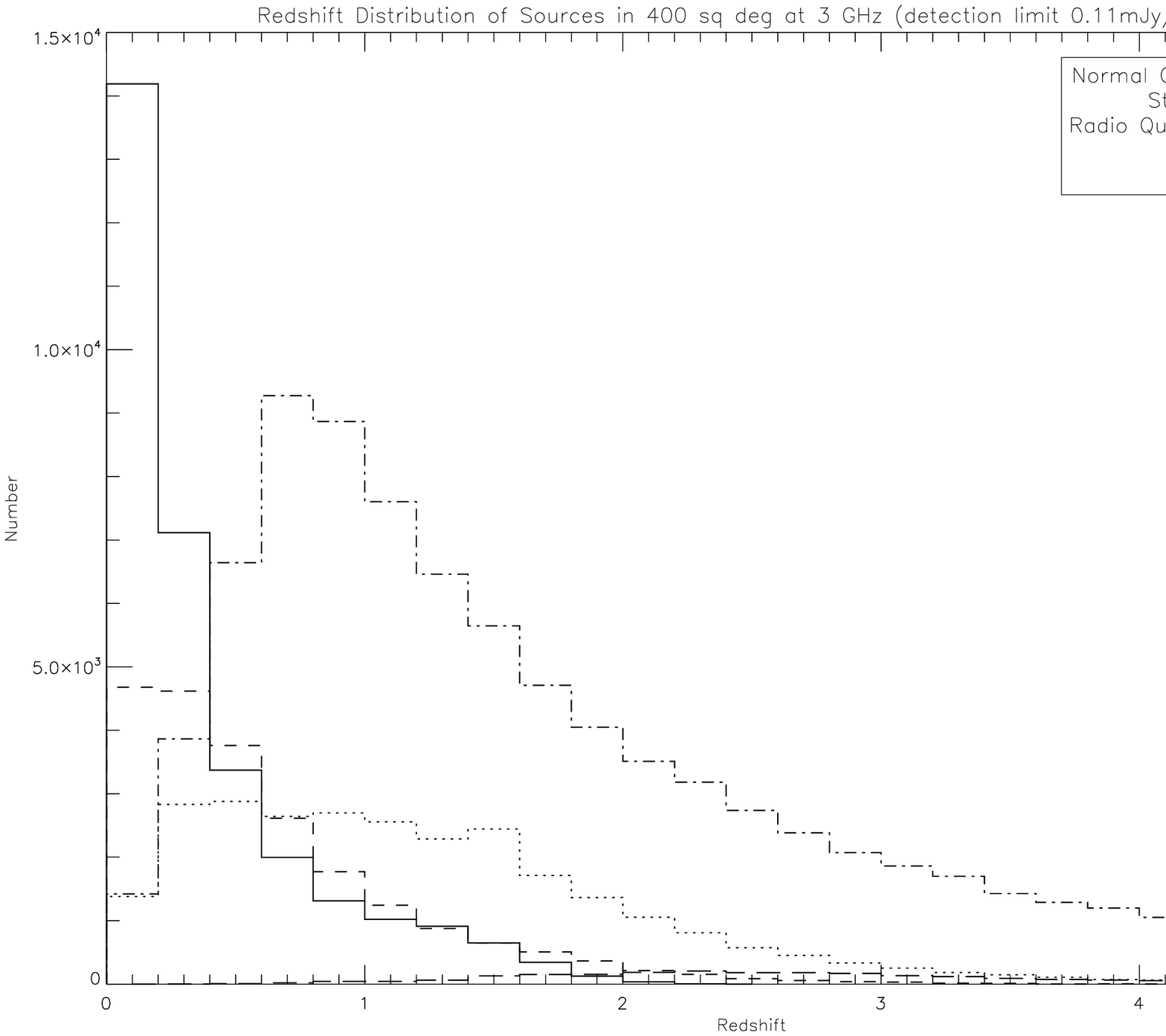}}%{\includegraphics[width=3in]{combined_histo_medium.eps}}\\

\noindent 
\hspace*{-2cm}
\subfloat[Deep]{\plotone{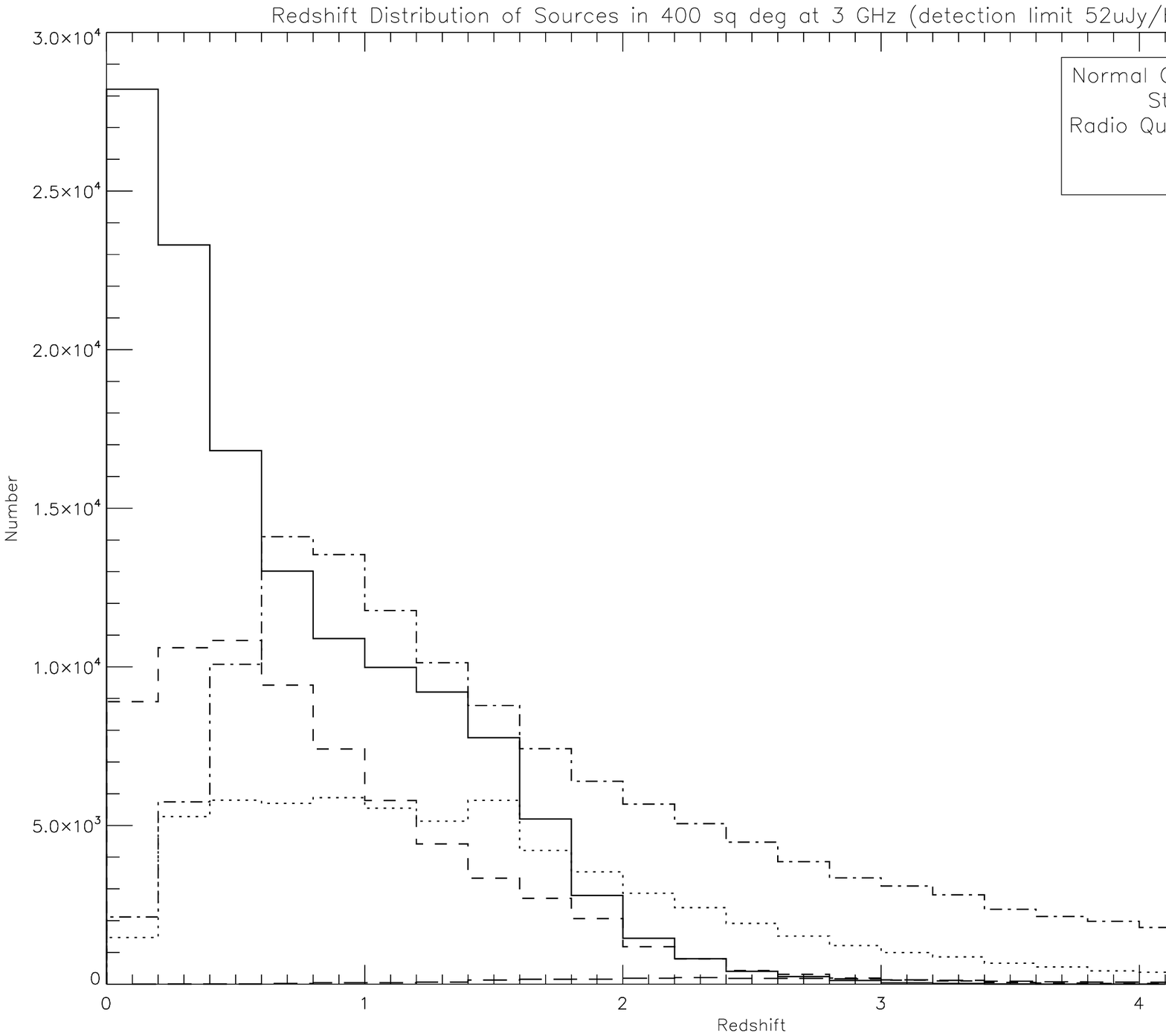}}%\includegraphics[width=3in]{combined_histo_deep.eps}}
\subfloat[Ultra-deep]{\plotone{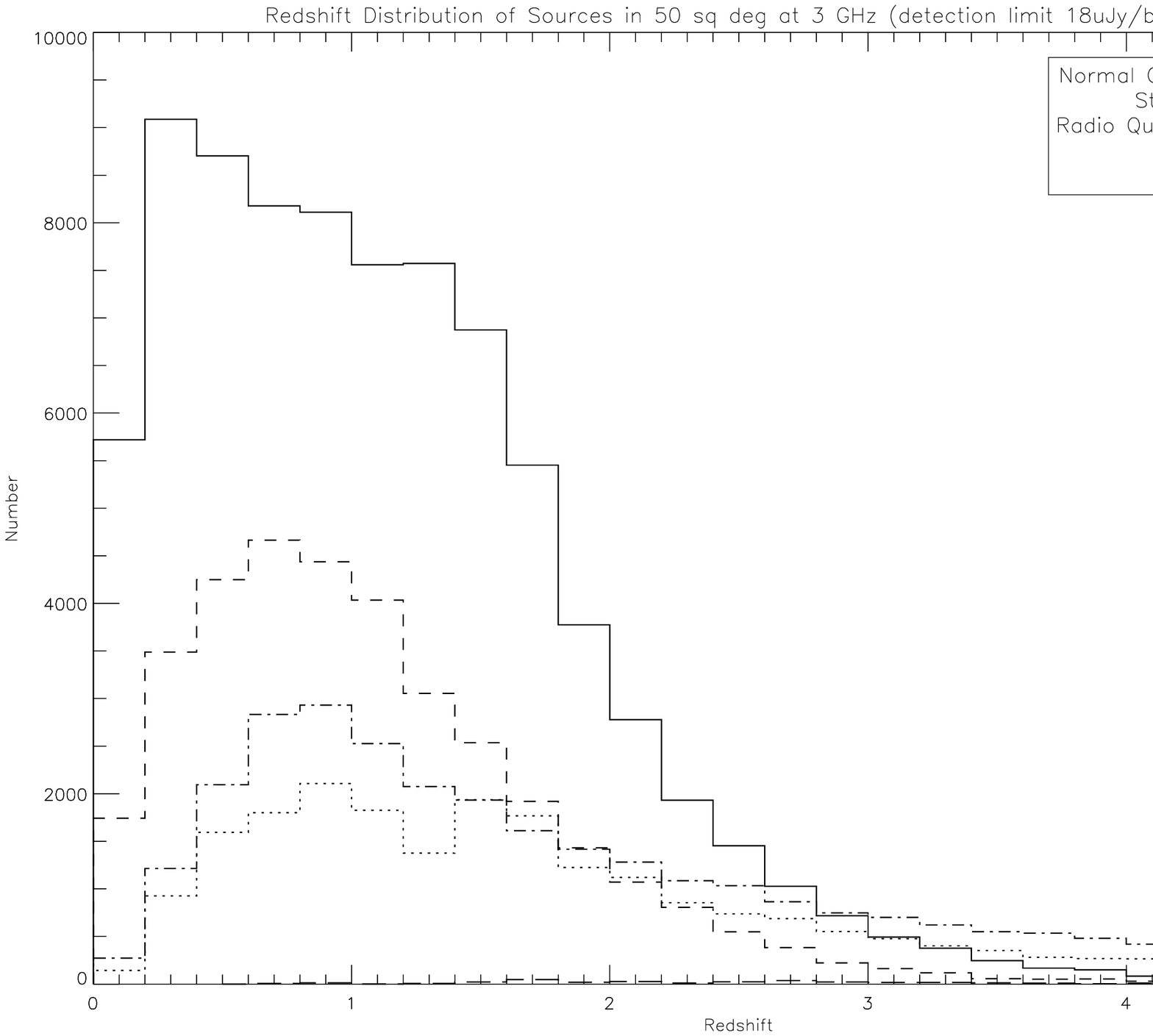}}%\includegraphics[width=3in]{combined_histo_ultradeep.eps}}

\caption[caption]{Expected redshift distribution of total intensity sources for a (a) shallow all-sky, (b) medium, (c) deep field and (d) ultra-deep field survey.} 
\end{center}

\end{figure}

\begin{figure}[h]
\centering
\plotone{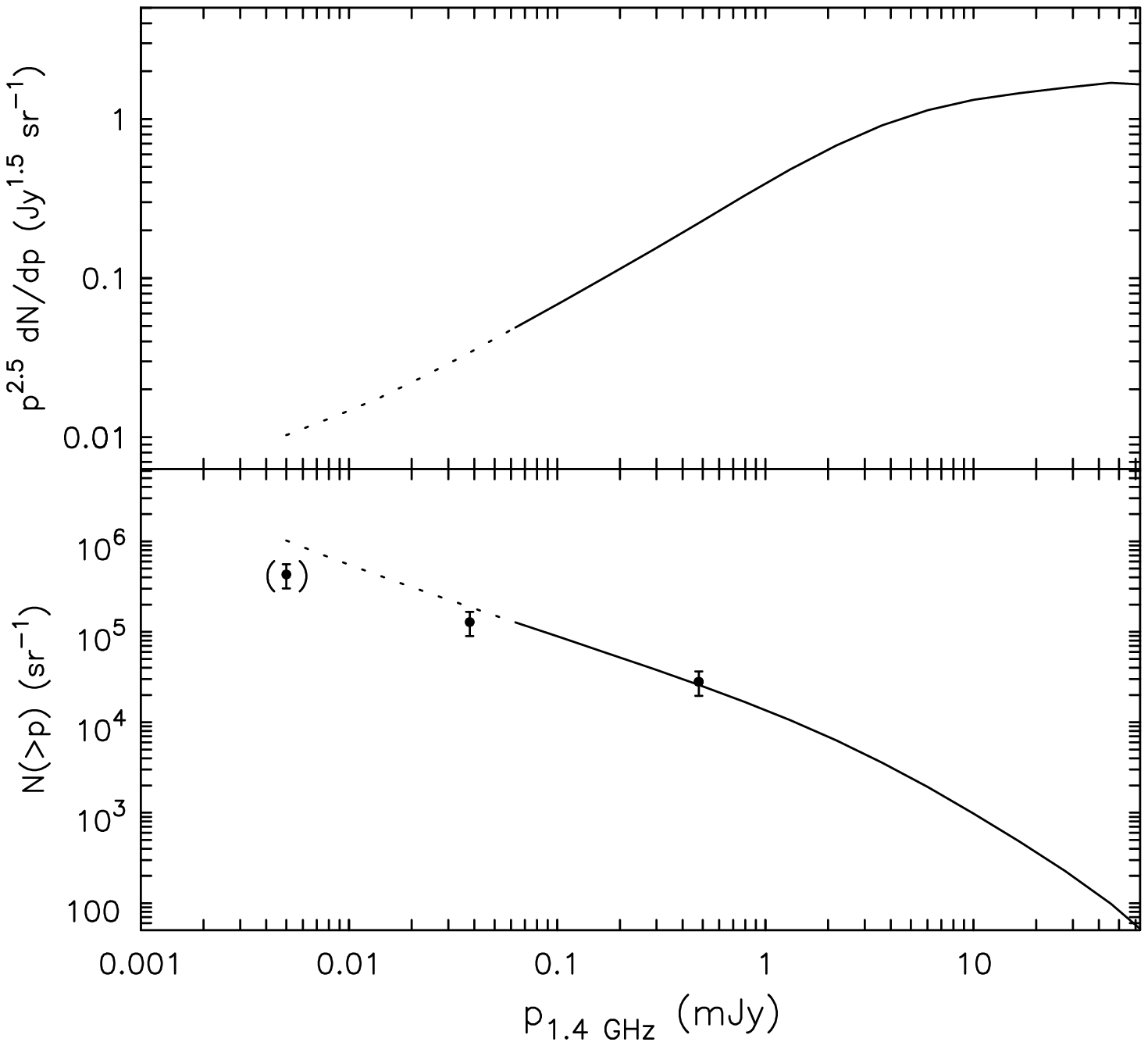}
\caption{The polarized source count (top panel) and the cumulative polarized source count (bottom panel) at 1.4 GHz derived from stacking the NVSS data (Stil et al. submitted). Source counts down to a flux density of 50 $\mu$Jy (solid line) are constrained. The source count below 50 $\mu$Jy is an extrapolation (dotted line). The cumulative source count extracted directly from deep observations of the GOODS North field at 1.4 GHz  (Rudnick \& Owen, submitted) has been overlaid as well.}
\end{figure}

\subsection{On and Off-axis Polarization Calibration}

In addition to flux density and complex gain calibrations, the proposed VLASS requires one to calibrate for the polarization angle and the instrumental leakages. The Jansky VLA is well-positioned for a polarization survey -- the on-axis leakage at S band is $\le$ 8 \% (Perley \& Haywood 2011) and most importantly, is stable on long term (Sault \& Perley 2009). To achieve high polarization fidelity across a wide field, off-axis polarization beam pattern needs to be corrected. It has been demonstrated\footnotemark[4]\footnotetext[4]{http://www.ira.inaf.it/meetings/rm2010/talks/Cotton.pdf}, using holographic observations at the Jansky VLA, that off-axis instrumental polarization can be well constrained and corrected for.

\section{Major Science Goals} 

% why S band
% Absoprtion line systems from SDSS up to X pairs.
%Any would do that . State which type (shallow, medium or deep) observations can allow us to understand this.
%First 5.4", astronmetry< 0.1" rms sensitivity~ 0.15mJy lower probably of chance alignment find more polarized counterparts
%NVSS 45" beam 2.5 mJy~rms sensitivity 

%circumgalactic medium.
%Britt's 2013 paper introduction is good - Mg II absorbers probe 10^4 K gas around galaxies, could probe SF outflow or cold mode acreeition. RM information will put limit on the level of ionization 
%ow can B field help to tell the difference? explained by expelled magnetized gas from the host galaxy or more pristine gas that is not %magnetized??? cold mode accretion??? Turn this around. How can B field help to solve the problem?
 %Scale of turbulence and value of |RM| could potentially help to distinguish resolve... 
%higher resolution and higher sensitivity, match quasar with radio counterparts reliably
%desirable 
%Amy Kimball work.higher resolution, better astronometry
%especially the SDSS area
\subsection{Properties of the Magneto-ionic Medium in Absorption Line Systems}
{\bf \noindent A wide-band polarization survey at $>$ 2 GHz is necessary to uniquely probe properties of ionized gas and magnetic fields associated with absorption line systems by modeling their Faraday depths and depolarization trends.}

Absorption line systems seen against polarized background sources enable us to study the magneto-ionic medium in these otherwise invisible systems. In particular, Mg II absorbers are associated with $\sim$10$^4$ K photo-ionized circum-galactic medium in a wide range of host galaxy types and redshifts (see e.g., Churchill et al. 2005 for a review). It is thought that they could potentially trace outflows from star formation (e.g., Norman et al. 1996) and cold-mode accretion (e.g., Kacprzak et al. 2010). 

The Faraday depth associated with the absorber provide a direct measure of the electron density and the magnetic field strength, both of which have little observational constrains. Observational limits on the electron density associated with Mg II clouds allow us to test photo-ionization models; while limits on the magnetic field strength enable us to infer the evolution of large-scale magnetic fields over cosmic time. For example, Bernet et al. (2008, 2013) interpreted the large $|$RM$|$ associated with Mg II absorbers as an evidence for the existence of $\mu$G field at z$\sim$1, possibly associated with outflows.

Recently, Bernet et al. (2012) have suggested that the observed Faraday rotation along sight lines towards Mg II absorbers is wavelength-dependent. The authors attributed this to depolarization caused by turbulence associated with the absorbing cloud. This has two important consequences: (1) It is necessary to observe at short wavelengths ($>$ 2 GHz) to ensure that the measured Faraday depth reflects that produced in the absorption line system. This is supported by the recent work of Joshi \& Chand (2012) who demonstrated, using 1.4 GHz NVSS RMs (Taylor et al. 2009) that Mg II absorbers produce non-zero Faraday depths only at $\sim$ 2$\sigma$ level; (2) Wide-band modeling of the depolarization trend of the background source can yield quantities such as the covering factor, and the characteristic turbulent cell size associated with the Mg II absorber, which could potentially be used to distinguish the origin of the absorbing gas. 

The most effective approach to significantly increase the number of polarized radio source-quasar pair is by conducting a deep and high angular resolution survey (to ensure good astrometry) overlapping with the SDSS sky coverage. We will utilize Mg II catalogs derived from the SDSS data (Quider et al. 2010, Zhu \& Menard 2013) to identify sight lines through absorbers (current total number $\sim$ 40,000). The wide bandwidth and the chosen high frequency at 2-4 GHz will allow one to model the Faraday structure through the absorbers and shed new light on the density, magnetic field strength, turbulent properties and covering fraction of this absorption line population.

\subsection{Thermal environments of AGNs and Radio Galaxies}
{\bf \noindent A wideband polarimetry survey provides a unique and independent way to quantify the thermal gas content in radio galaxies, allowing us to better understand AGN feedback.}

Feedback from active galactic nuclei are important in galaxy formation: they could suppress cooling in massive galaxies, producing the bright-end cut-off of the luminosity function (e.g. Best 2006, Croton 2006). AGN activities appear to be intimately linked to the star formation history because the peak in cosmic star formation coincides with that for AGN activities (e.g., Hopkins \& Beacom 2006). The nature of this AGN feedback is very much under debate: it has been shown that energy deposited by radio jets can either trigger or quench star formation (e.g., Wagner et al., 2012). AGNs are also thought to influence their surrounding intergalactic medium by enrichment of metals (Aguirre et al. 2001) and magnetic fields (Furlanetto \& Loeb 2001). Thus, investigating how radio galaxies impart energies into the ISM/IGM is crucial in understanding their impacts on galaxy and structure formation. 

While minimal interaction between radio lobes and the environment would lead to a thin ``skin" of thermal material around the lobes (e.g., Bicknell et al. 1990), significant interaction would lead to large-scale mixing of thermal gas with the synchrotron emitting material throughout the lobe, causing internal Faraday dispersion. Modeling of the depolarization trend using wide-band polarization data is the key to distinguish between these two scenarios. O'Sullivan et al. (2013) fitted the depolarization trend of the lobes of Centaurus A and found a thermal gas of density 10$^{-4}$ cm$^{-3}$ well mixed in with the synchrotron emitting gas in the lobes. This provides an independent way to estimate the internal lobe pressure contributed by thermal gas, and to determine the mass of the material entrained in the flow. A wideband polarimetry survey allows us to quantify the thermal gas content in radio galaxies across luminosities, redshifts and environments, which will enable us to understand the physics behind how these lobes are driven, and hence on AGN feedback in general.

\subsection{Evolution of AGNs and Galaxies over Cosmic Time}
{\bf \noindent Wideband polarization information of extragalactic sources allow us to derive a k-correction to shift the polarization measurements to the rest-frame of the emitter, and thus it facilitates the comparison of source properties as a function of z.}

We now have exciting extragalactic source catalogs of rotation measure-vs-z (e.g., Hammond et al. 2012). However, we are severely limited by our inability to correct polarization data for cosmological expansion. These k-corrections are necessary for fractional polarization and Faraday depths. Provided that we have the redshift of the source from spectroscopy (or photometric z), a ``polarization k-correction" is only possible with wideband polarized spectral energy distribution. Once the polarization measurements are shifted into the rest-frame of the emitter, we can infer intrinsic properties of the magneto-ionic medium in AGNs, galaxies and their immediate environments and investigate the evolution of their rest-frame properties as a function of redshift. %(May be add Klaus's figure here)

\subsection{Resolve the Mystery of Polarized Source Counts at the Low Flux Density End}
{\bf \noindent A wideband spectro-polarimetric survey of the extragalactic sky from 2-4 GHz will allow us to test the origin of the observed increase in percent polarization with decreasing flux densities.}

%provides the necessary data to include with other wavelengths (i.e., optical, infrared) to understand the formation and evolution of galaxies across cosmic time.}
Faint radio sources appear to be more highly polarized in the NVSS catalogue (Mesa et al. 2002, Tucci et al. 2004). The same result was reported for the ELAIS N1 field by Taylor et al. (2007) and Grant et al. (2010), and for the ATLBS sources by Subrahmanyan et al. (2010). However, Hales et al. (2013) have attributed this entirely to selection effects, which cast doubt on earlier results that had found such a dependence. 

Recently, Banfield et al. (2013, submitted) show that the correlation between percentage polarization and flux density depends on the host galaxy of the radio emission and the environment that the radio galaxy is in. Cross-matching the radio sources with their infrared counterparts from WISE, the authors found that radio AGNs that live in the most luminous infrared galaxies tend to have lower percentage polarization at 1.4 GHz, larger linear size, and are consistent with High-redshift Radio Galaxies (HzRGs) than radio AGNs that live in quiescent galaxies. The lower percentage polarization in dusty AGNs at high z is thought to be due to internal Faraday dispersion within their lobes, where the large-scale field is tangled and the synchrotron emitting and Faraday rotating material is mixed in. The proposed VLASS can significantly increase the sample size at the low flux density end to help resolve this long standing mystery. Furthermore, the wide-band polarimetric data will allow one to model the depolarization associated with individual sources and to test the internal Faraday dispersion hypothesis suggested in Banfield et al. (2013). %We note that a shallow all-sky field (detection threshold $\sim$ 0.45 mJy) may not be sensitive enough for this particular science goal. 

\subsection{Properties of Magnetic Fields in Galaxies}
{\bf \noindent A deep S band polarization survey is well suited for a statistical study of magnetic fields in unresolved normal and starburst galaxies.} 

%largest scale sensitive to is $\sim$ 8' which is sufficient in getting some galaxies in the local volume!
Imaging the spatially resolved polarized synchrotron emission from galaxies in the local universe show the existence of $\mu$G azimuthal fields (e.g., Beck et al. 1996). The evolution of galactic-scale magnetic fields is poorly constrained as this traditional approach becomes increasingly challenging for distant galaxies. An alternative approach is to utilize the statistics of integrated synchrotron polarization of unresolved galaxies to infer their general magnetic field properties (e.g., Stil et al. 2009). Star-forming normal and starburst galaxies become an increasingly important fraction of all radio sources at low flux densities, thus a deep survey is needed to recover them. 

Using models of magnetic fields in Milky-Way type galaxies, Stil et al. 2009 and Sun \& Reich (2012) have showed that the average fractional polarization of unresolved Milky-Way type galaxies is a factor of 3-4 higher at frequencies $>$ 2 GHz than at 1.4 GHz. Moreover, Braun et al. (2010), using the SINGS galaxy sample, suggested that only halo magnetic fields can be probed at L band, as the galactic disk is completely depolarized at this frequency. It is only when the observation is carried out at higher frequencies that one is sensitive to both the disk and halo magnetic fields. On the other hand, strong turbulence in starbursts would lead to severe Faraday depolarization at long wavelengths, and hence they may only be polarized at higher frequencies. Depolarization trends of unresolved galaxies encode information on the turbulent small-scale fields. Therefore, a wide band polarization survey at 2-4 GHz is necessary to statistically study both the large-scale and small-scale magnetic fields in unresolved star forming galaxies. Modeling the probability distribution function of polarization of unresolved galaxies at a range of z can reveal any systematic changes in magnetic field scales and properties over cosmic time. These results can be used to test against theoretical predictions from various galactic magnetic field generation mechanisms (see for example, Heiles \& Zweibel 1997) to constrain the poorly measured galactic magnetic field amplification time scale and the seed field strength (Arshakian et al. 2009). We note that a shallow all-sky survey is unlikely to be deep enough for this science goal (the redshift range of normal galaxies detected in total intensity is $<$0.07). A deep field targeting a galaxy over density region (such as part of the Sloan Great Wall) is preferred.

%\section{Ancillary data/ partner surveys at other wavelengths} 
%POSSUM,
%WONDAN,
%SDSS / any other spectroscopic redshift surveys,
%IR surveys
%Any other upcoming surveys? 

\section{Extraction of Polarization Information} 

We will take advantage of the pipeline developed for the POSSUM project to process polarization data from the proposed VLASS. Assuming the total intensity and polarization data have been fully calibrated (both on and off-axis effects) we create a catalogue of Stokes I emission using the Aegean source finder. Aegean is a next generation source finder that reliably decomposes multiple overlapping sources into Gaussian components (Hancock et at 2012). The POSSUM pipeline measures the polarized properties of each compact Stokes I detection. For each source the Stokes I, Q, U and V spectra are extracted and the local noise level estimated using robust statistics. The pipeline performs RM-synthesis using the Q/I and U/I spectra to negate spectral index effects, resulting in a `dirty' Faraday dispersion function (FDF): the FDF convolved with the frequency-sampling window, called the RM spread function (RMSF). The complexity of the FDF is quantified by deconvolving using the CLEAN algorithm and the polarized intensity, polarized fraction, RM and uncertainties (or limits in the case of non-detections) are measured from the peak. Sources with complex FDFs require further analysis to disentangle multiple-peaked FDFs. In these cases representative models of Faraday thick screens are fit to the Q/I and U/I spectra using a maximum-likelihood method. Initial parameters estimated from RM-synthesis allow the fitting to converge rapidly.

The final product of the POSSUM pipeline is a catalogue of compact polarized sources with fully characterized polarization properties. The POSSUM team have also developed a web-based visualization interface to the catalogue. Subsets of of the catalogue can be selected and visualized using a structured query language (SQL) interface allowing rapid quality control and analysis. 

\section{References}
\scriptsize{
\noindent Aguirre, A., Hernquist, L., Schaye, J., et al. 2001, ApJ, 561, 521 \\
Arshakian, T. G., Beck, R., Krause, M., \& Sokoloff, D. 2009, A\&A, 494, 21 \\
Arshakian, T. G., \& Beck, R. 2011, MNRAS, 418, 2336 \\
Banfield et al. 2013, submitted \\
Beck, R., Brandenburg, A., Moss, D., Shukurov, A., \& Sokoloff, D.1996, ARA\&A, 34, 155 \\
Bernet, M. L., Miniati, F., Lilly, S. J., Kronberg, P. P., \& Dessauges-Zavadsky,M. 2008, Nature, 454, 302 \\
Bernet, M. L., Miniati, F., \& Lilly, S. J. 2012, ApJ, 761, 144 \\
Bernet, M. L., Miniati, F., \& Lilly, S. J. 2013, ApJ, 772, L28 \\
Best, P. N., Kaiser, C. R., Heckman, T. M., \& Kauffmann, G. 2006, MNRAS, 368, L67 \\
Bicknell G.V., Cameron R.A. \& Gingold R.A. 1990, ApJ, 357, 373 \\
Bower G. C., Falcke H., Mellon R.R., 2002 ApJ, 578, L103 \\
Braun, R., Heald, G., \& Beck, R. 2010, A\&A, 514, A42 \\
Brentjens M. A., de Bruyn A. G., 2005, A\&A, 441, 1217 \\
Burn B. J., 1966, MNRAS, 133, 67 \\
Churchill, C. W., Kacprzak. G. G., \& Steidel. C. C. 2005, IAU Colloquium No. 199 \\
Croton, D. J., Springel, V., White, S. D. M., et al. 2006, MNRAS, 365, 11 \\
Farnsworth, D., Rudnick, L., \& Brown, S. 2011, AJ, 141, 191 \\
Furlanetto, S. R., \& Loeb, A. 2001, ApJ, 556, 619 \\
Gaensler, B. M., Haverkorn, M., Staveley-Smith, L., et al. 2005, Science, 307, 1610 \\
Gaensler, B. M., Landecker, T. L., Taylor, A. R. \& POSSUM Col- laboration, 2010, BAAS, 42, 515 \\
Goodlet, J. A. \& Kaiser, C. R. 2005, MNRAS, 359, 1456 \\
Grant J.K., Taylor A.R., Stil J.M. et al. 2010, ApJ, 714, 1689 \\
Hales et al. submitted \\
Hammond A. M., Robishaw T., Gaensler B. M., 2012, preprint (arXiv:1209.1438) \\
Hancock, P. J., Murphy, T., Gaensler, B. M., Hopkins, A., Curran, J. R 2012, Astrophysics Source Code Library, record ascl:1212.009 \\
Heald, G., Braun, R., \& Edmonds, R. 2009, A\&A, 503, 409 \\
Hopkins A.M. \& Beacom J.F. 2006, ApJ, 651, 142 \\
Kacprzak, G. G., Churchill, C. W., \& Nielsen, N. M. 2012, arXiv:1205.0245 \\
Joshi, R. \& Chand, H. 2013, MNRAS, 434, 3566 \\
Mao, S. A., Gaensler, B. M., Haverkorn, M., et al. 2010, ApJ, 714, 1170 \\
Mesa D., Baccigalupi C., De Zotti G. et al. 2002, A\&A, 396, 463 \\
Norman, C. A., Bowen, D. V., Heckman, T., Blades, C., \& Danly, L. 1996, ApJ, 472, 73 \\
OÕSullivan S. P., et al., 2012, MNRAS, 421, 3300. \\
OÕSullivan, S. P., Feain, I. J., McClure-Griffiths, N. M., et al. 2013, ApJ, 764,162 \\
Perley, R. \& Hayward, B. 2011, EVLA memo 151 \\
Quider, A. M., Nestor, D. B., Turnshek, D. A., et al. 2011, AJ, 141, 137 \\
R\"ottgering, H., et al. 2010, http://www.astron.nl/radio-observatory/APERTIF-eoi-abstracts-and-contact-information \\
Rudnick \& Owen, submitted  \\
Sault, R.J.  \& Perley, R. 2009, EVLA memo 135 \\
Schnitzeler, D. H. F. M., Katgert, P. \& de Bruyn, A. G. 2009, A\&A, 494, 611 \\
Sokoloff D. D., Bykov A. A., Shukurov A., Berkhuijsen E. M., Beck R., Poezd A. D., 1998, MNRAS, 299, 189 \\
Stil, J. M., \& Taylor, A. R. 2007, ApJ, 663, L21 \\
Stil, J. M., Krause, M., Beck, R., \& Taylor, A. R. 2009, ApJ, 693, 1392 \\
Stil, Keller, George \& Taylor, submitted  \\
Subrahmanyan R. et al. 2010, MNRAS, 402, 2792 \\ 
Sun, X. H. \& Reich, W. 2012, A\&A, 543, A127 \\
Taylor A.R., Stil J.M., Grant J.K. et al. 2007, ApJ, 666, 201 \\
Taylor A. R., Stil J. M., Sunstrum C., 2009, ApJ, 702, 1230 \\
Tribble P. C., 1991, MNRAS, 250, 726 \\
Tucci, M., Martinez-Gonzalez, E., Toffolatti, L., Gonzalez-Nuevo, J., \& De Zotti, G. 2004, MNRAS, 349, 1267 \\
Wagner A.Y., Bicknell G.V. \& Umemura M. 2012, ApJ, 757, 136 \\
Wilman, R. J., et al. 2008, MNRAS, 388, 1335 \\
Wilman, R. J., Jarvis, M. J., Mauch, T., Rawlings, S., \& Hickey, S. 2010, MNRAS, 405, 447 \\
Zhu, G. \& Menard, B. 2013, ApJ, 770, 130 \\
Zweibel, E. \& Heiles, C. 1997, Nature, 385, 131 \\
}
\end{document}